# Embedding Sustainability in Complex Projects: A Pedagogic Practice Simulation Approach

Caroline N J Tite [a,*], David Pontin [a], & Nicholas Dacre [b]

[a] Warwick Manufacturing Group, University of Warwick, Coventry, CV4 7AL, UK
[b] University of Southampton Business School, University of Southampton, Southampton, SO17 1BJ, UK
[*] Corresponding Author: caroline.tite@warwick.ac.uk

## Abstract

Sustainability is focussed on avoiding the long-term depletion of natural resources. Under the terms of a government plan to tackle climate change, a driver for improved sustainability is the cut of greenhouse gas emissions in the UK to almost zero by 2050. With this type of change, new themes are continuously being developed which drive complex projects, such as the development of new power generation methods, which encompass challenging lead times and demanding requirements. Consideration of the implementation of strategies and key concepts, which may engender sustainability within complex projects therefore presents an opportunity for further critical debate, review, and application through a project management lens. Sustainability incorporation in project management has been documented in academic literature, with this emerging field providing new challenges. For example, project management education can provide a holistic base for the inculcation of sustainability factors to a range of industries, including complex projects. Likewise, practitioner interest and approaches to sustainability in project management are being driven by the recently Chartered Association for Project Management (APM). Whilst this body makes a significant contribution to the UK economy across many sectors, it also addresses ongoing sustainability challenges. Therefore, by drawing on research and practitioner developments, the authors argue that by connecting with the next generation through practice simulation approaches, and embedding sustainability issues within project management tools and methods, improved focus on sustainability in complex project management may be achieved.

**Keywords**: Complex Project Management, Sustainability, Sustainable Projects, Pedagogy, Project Management, Climate Change, Practice Simulation, Association for Project Management, APM.

## Introduction

There is widespread agreement that sustainability is an area of focus which needs to be considered and addressed within responsible management practice (Gkogkidis & Dacre, 2021). Many authors address this subject area and consider methods for embedding sustainability within a project management curriculum. For example, the Association for Project Management (APM) recognise the need for embedding sustainability at the project management level, although there is a wider global organisational perspective which also merits a high degree of contemplation. In considering an approach towards this aim, it should be underpinned by a clear definition of what is meant by sustainability in project management education, with an agreed framework which can be implemented as a reference point in formalising the context in providing a robust and rigorous syllabus. For example, Glasser and Hirsh (2016) argue for a set of sustainability core competencies, in order to develop robust learning for Sustainability. Further research is required in order to identify these competencies, towards developing a framework for implementation which can be embedded in higher education.

Thürer et al. (2018) notes that there is general, but not complete agreement on three dimensions to sustainability and sustainable development, which include: (i) Social, (ii) Economic and (iii) Environmental. Segalàs et al. (2010) wrote of a fourth dimension, being the (iv) Institutional Dimension, which encompasses the role of education and external stakeholders. While Watson et al. (2013, p. 106) argue that there has been a "rapid increase on the number of engineering schools in higher education institutions that have incorporated sustainability into their teaching". Lambrechts et al. (2013) note that sustainability appears to be integrated in a peace-meal fashion.

From their review of the literature Thürer et al. (2018, p. 616) found what appeared to be "a lack of studies on implementations in developing economies", and they called for more studies of how to grow and engender sustainable development in Europe and the US in particular.

Therefore, a set of core competencies in sustainability needs to be identified, analysed, and developed in order to guide degree course and curriculum development (Dacre et al., 2019; Reynolds & Dacre, 2019), which may also guide changes in approach to sustainability education within academic institutions. Complex project management may be best placed as a basis for determination of relevant concepts forming the basis of this set of competencies. This research paper thus focuses on framing the concept of sustainability by drawing on project management literature, ensued by a practice simulation vignette stemming from extensive academic insights across different Higher Education Institutions (HEIs).

## Framing Sustainability

Experts in education for sustainable development consider that institutional and social aspects are generally more relevant to sustainability than environmental aspects (Segalàs et al., 2012). This is in clear contrast to students and potentially the wider public, which mostly perceive the environmental aspect to be at the centre of sustainability and sustainable development. For example, Kagawa (2007) survey among students at Plymouth University found that almost half of the respondents related sustainability and sustainable development primarily with the environment while social, economic, political, and cultural dimensions of sustainability were less represented and remained marginal in the understanding of most students.

Redressing this balance in the favour of social and institutional aspects is consequently seen by many researchers as a key task of education for sustainable development (Boks & Diehl, 2006; Kagawa, 2007; Segalàs et al., 2010; Segalàs et al., 2012). Thürer et al. (2018, p. 609) asked "what is the current state-of-the-art on integrating sustainability and sustainable development into engineering curricula?"

The degree of change in the curricula ranges from new material on sustainability in an existing module, to a new module on sustainability in an existing program, to an entirely new program of study on sustainability. Based upon this systematic review, twelve important future research questions emerged. This includes the exploration of the knowledge and value frameworks of students and teachers, the exploration of stakeholder influences, including accreditation institutions, industry partners, parents and society, and the use of competencies for the evaluation of implementations.

There is no doubt that there is a strong political will and commitment towards sustainability and sustainable development (Thürer et al., 2018). University leaders and educators have begun to recognise the importance of sustainability and sustainable development.

There is a growing body of literature on the integration of sustainability and sustainable development into the curricula at universities around the globe. However, to the question – What should we learn on sustainable development? – Svanström et al. (2012) answer was: (i) what are the problems; and (ii) how should they be solved. During this review, it was felt that most of the cases focused on creating environmental awareness and system thinking when identifying problems and solving them.

There is a plethora of approaches to embedding sustainability within the higher education context. Haney et al. (2020) reported that developing sustainability leaders requires not just new knowledge and skills but also new ways of thinking and ultimately an underlying motivation to act.

Making sustainability personal for individual leaders is a key aspect for embedding sustainability and this can be achieved through learning outcomes. A framework creates a foundation for individual leaders to gain personal understanding, a feeling of being committed to the required change and through these, empowerment to act in embedding sustainability in their organisations.

Personal dimensions and a softer, more individual dimension of sustainability competencies have been previously illustrated as being particularly salient. Thus, framing sustainability and sustainable development is key to understanding the core concepts which are required in a project management curriculum. As such, in this research the authors also include sustainability core competencies specific to this area.

Bringing these ideas and influences together within a complex project management area are important, and leadership aspects are key in understanding at a personal level. It is also important in relation to the area of complex project management because of the area of leadership which brings a range of disciplines together in order to achieve successful project outcomes.

## Practice Simulation

The authors' experience of engaging diverse groups across different HEIs suggests that many individuals are well informed and highly motivated towards projects with a strong sustainability element. When participants are presented with choice over complex projects to research, many select those with sustainability elements or associated ethical elements, such as green buildings, renewable energy, electric vehicles and associated infrastructure, CSR projects, healthcare in developing countries, and charity sector projects.

This suggests that rather than just raising awareness of sustainability within complex projects, effort should also be made on defining the particular skills required for effective sustainable project delivery, including developing practice simulation exercises where these can be reviewed, developed and crafted. Practice simulation exercises have become a popular teaching approach in project management courses as they help foster a conducive learning environment where participants can apply theory and current practice (Dacre et al., 2019; Pontin & Adigun, 2019; Yao & Tite, 2019).

They provide an opportunity to combine hard or technical project management skills such as scope definition and work de-construction using Work Breakdown Structure (WBS), planning, scheduling and cost estimating with soft or interpersonal skills, such as effective communication, leadership, motivation, influencing and problem solving within a multi-disciplinary team setting (Dacre et al., 2019; Gkogkidis & Dacre, 2020).

A carefully developed exercise can simulate not only the planning phase but also the execution phase and can introduce learners to the dynamic and often disordered nature of real-life projects (Gkogkidis & Dacre, 2021; Pontin & Adigun, 2019; Yao & Tite, 2019).

The ability to effectively communicate, motivate and influence others, have difficult conversations, work across different cultures, reflect and be self-aware are some of the qualities business leaders suggest are required in employees but do not always have the level desired (CMI, 2018).

Practice simulations exercises provide an environment where these skills can be tested and developed (Pontin & Adigun, 2019; Yao & Tite, 2019) but without the commercial consequences if mistakes are made in the learning journey. Researchers support the role of a gamified 'play-based approach' as part of an innovative learning environment in HEIs (Dacre et al., 2015; Smith, 2019). However, there are gaps. Many of the simulation exercises cover the planning phase, but less so the execution phase and often risk management is comprehensively embedded (Zwikael & Gonen, 2007).

Where the execution phase and the possibility of risk events occurring are introduced, the practice simulation becomes closer to real-life and provides a more accurate and detailed educational experience. Such exercises enjoy excellent learner engagement, positive feedback and further help develop those skills that employers require to support he growth of their organisations (Dacre et al., 2018; Pontin & Adigun, 2019). However, further research is required to assess the transference of these skills and whether there is evidence that these can be adapted in the workplace (Romero et al., 2015).

TEDI-London is a design-led engineering school that is taking the concept of project-based practice simulation to a higher level with up to 55% of the learning experience based on project or scenario-based approaches. Founded by 3 universities with strong research credentials, the pedagogy is described as 'disruptive' and aims to transform education by providing new qualifications at graduate and post graduate level in Global Design Engineering (Raper, 2019). In this vein of embedding practice simulation into pedagogical approaches for complex project management, the authors are also developing new research informed teaching materials with a heavier emphasis on project management applications.

A new set of modules aim to provide learners with a systematic understanding of knowledge of how to apply project management tools and methods within a range of different industries and sectors. It is the hope that participants will gain a critical awareness of challenges of application of a range of project tools and methods in the areas including, business change project management, engineering project management, IT project management, and construction project management.

This will enable participants to consider and establish new insights in application of project management methods, being mindful of advances in this area of both academic study and related professional practice related to established project management methodologies, including a focus on sustainability.

Assessment for these new sets of modules will focus of the quality of the proposed project management applications and learners will need to demonstrate the particular constraints and peculiarity or the requirements of the industry or sector chosen. Originality, creativity, innovation or problem solution (new solution or transference of a solution from another industry or sector), will be valued and assessed along with the reasoning and justification behind the proposed solution.

Effective communication of the proposed solution to a simulated executive audience will also test and allow development of this important skill for project managers. All these skills are equally important for effective sustainable project delivery. This type of module design where new sectors, and project management development can be added easily and progressively within the module and curriculum design, lends itself to overarching sustainability themes.

Stakeholder management or more recently stakeholder engagement is a recognised knowledge area within project management professional associations' body of knowledge, and the interest and evolution of critical insights over the last decade has developed in line with thinking on front ending (project selection techniques), and back ending (benefits realisation).

Closely linked with the soft or interpersonal skills, effective stakeholder engagement draws on deep listening and communication skills, strong leadership coupled with hard or technical stakeholder analysis skills. In a world where priorities are in transition, such a skill set is particularly relevant for successful project delivery where an enhanced sustainability element exists.

Adding a stakeholder engagement element to the project management practice simulation could then enrich the learning experience further and test, assess, and allow learning related to this important emerging knowledge area. Within such developments in engineering and project management teaching and learning, community elements of sustainability should be embedded.

A balanced curriculum including case study examples covering both successful and failed projects with strong sustainability elements, pedagogy around the skillset required and an opportunity to apply, practice, and develop through project simulation exercises could offer insights in embedding sustainability into complex project management pedagogy (Gkogkidis & Dacre, 2020, 2021).

Essentially, the authors do not claim to have achieved all of these aims, but rather illustrate how critical thinking is developing. This paper therefore serves to invite critical discourse around other ideas and potential collaboration as the pedagogic project management community seek to embed sustainability into programmes with benefits for learners and wider society.

## Discussion

For project management the definition of sustainability and their understanding of aspects of sustainability are important. Rampasso et al. (2019) found that according to the learners' perception, the reason for the eliminations of Concern With Employees (CWE); Support for Local Communities (SLC); is related to their perceptions.

In this case, the participants from our sample did not consider these issues when analysing sustainability. That is, for them the parameters related to employees and local communities, which are not included in their sustainability analysis. The reason for this is due to the fact that learners generally were found to have low correlation with the overall sustainability grade, i.e. when participants evaluate sustainability, they are not considering those parameters. So, what is sustainability for these learners in the context of their development?

There are many examples of researches that point out the excessive focus on environmental sustainability as a problem in education (Björnberg et al., 2015; Guerra et al., 2017; Yuan & Zuo, 2013). In this research, learners do not consider this as a primary. This means that participants consider its parameters as the least important.

Another argument for project managers to bring a personal understanding and thereby impetus to the embedding of sustainability in engineering projects.

There is evidence of a divergence in the approach to embedding sustainability in different countries, cultures and courses. (Dagiliūtė et al., 2018) conducted research in two universities from Lithuania and the results pointed out that learners consider that social sustainability is the most important aspect of a sustainable institution.

In their research, respondents were a small part of their sample. It is clear that participants from the sample are better prepared in relation to social aspects of sustainability, especially the concerns regarding employees and local communities. This is particularly important when it is considered the role of project managers in the development and implementation of complex projects and production systems (Rampasso et al., 2019). The link between sustainability concerns and the development of new products and services is important but it is not regarded as being wholly sufficient.

Project managers concerned with embedding sustainability into their practices must present reasonable levels of concern regarding environmental issues, such as a sustainable use of water and energy, limiting emission of polluting gases, adhering to legislations, including acting responsibility in all aspects of sustainable project management.

Although it is an exploratory research, the findings in this paper can be useful for researchers as starting point for others studies and for professors and program coordinators from HEIs who can further elaborate on these findings to analyse their own project management programs, and evaluate priorities in the improvements they perform. Therefore, as a future research it is recommended to further develop this methodological procedure in other project management programs across a fuller range of HEIs in order to broaden the debate about perceptions on sustainability issues in complex projects.

## References

Björnberg, K. E., Skogh, I.-B., & Strömberg, E. (2015). Integrating social sustainability in engineering education at the KTH Royal Institute of Technology. International Journal of Sustainability in Higher Education.
https://doi.org/10.1108/IJSHE-01-2014-0010

Boks, C., & Diehl, J. C. (2006). Integration of sustainability in regular courses: experiences in industrial design engineering. Journal of Cleaner Production, 14(9-11), 932-939.
https://doi.org/10.1016/j.jclepro.2005.11.038

CMI. (2018). 21st Century Leaders: Building Employability Through Higher Education. Chartered Management Institute.
https://www.managers.org.uk/wp-content/uploads/2020/03/21st_Century_Leaders_CMI_Feb2018.pdf

Dacre, N., Constantinides, P., & Nandhakumar, J. (2015). *How to Motivate and Engage Generation Clash of Clans at Work? Emergent Properties of Business Gamification Elements in the Digital Economy*. International Gamification for Business Conference, Birmingham, UK.
https://dx.doi.org/10.2139/ssrn.3809398

Dacre, N., Gkogkidis, V., & Jenkins, P. (2018). *Co-Creation of Innovative Gamification Based Learning: A Case of Synchronous Partnership*. Society for Research into Higher Education, Newport, Wales UK. https://dx.doi.org/10.2139/ssrn.3486496

Dacre, N., Senyo, PK., & Reynolds, D. (2019). *Is an Engineering Project Management Degree Worth it? Developing Agile Digital Skills for Future Practice*. Engineering Education Research Network, Coventry, UK. https://dx.doi.org/10.2139/ssrn.3812764

Dagiliūtė, R., Liobikienė, G., & Minelgaitė, A. (2018). Sustainability at universities: Students' perceptions from Green and Non-Green universities. Journal of Cleaner Production, 181, 473-482. https://doi.org/10.1016/j.jclepro.2018.01.213

Gkogkidis, V., & Dacre, N. (2020). *Co-Creating Educational Project Management Board Games to Enhance Student Engagement*. European Conference on Game Based Learning, Brighton, UK. https://dx.doi.org/10.2139/ssrn.3812772

Gkogkidis, V., & Dacre, N. (2021). The Educator's LSP Journey: Creating Exploratory Learning Environments for Responsible Management Education Using Lego Serious Play. *Emerald Open Research*, *3*(2). https://doi.org/10.35241/emeraldopenres.14015.1

Glasser, H., & Hirsh, J. (2016). Toward the development of robust learning for sustainability core competencies. Sustainability: The Journal of Record, 9(3), 121-134.

Guerra, A., Ulseth, R., Jonhson, B., & Kolmos, A. (2017). Engineering grand challenges and the attributes of the global engineer: a literature review. 45TH SEFI ANNUAL CONFERENCEEuropean Society for Engineering Education. Annual Conference proceedings,

Haney, A. B., Pope, J., & Arden, Z. (2020). Making It Personal: Developing Sustainability Leaders in Business. Organization & Environment, 33(2), 155-174. https://doi.org/10.1177/1086026618806201

Kagawa, F. (2007). Dissonance in students' perceptions of sustainable development and sustainability. International Journal of Sustainability in Higher Education. https://doi.org/10.1108/14676370710817174

Lambrechts, W., Mulà, I., Ceulemans, K., Molderez, I., & Gaeremynck, V. (2013). The integration of competences for sustainable development in higher education: an analysis of bachelor programs in management. Journal of Cleaner Production, 48, 65-73. https://doi.org/10.1016/j.jclepro.2011.12.034

Økland, A. (2015). Gap analysis for incorporating sustainability in project management. Procedia Computer Science, 64, 103-109. https://doi.org/10.1016/j.procs.2015.08.469

Pontin, D., & Adigun, L. (2019). The value of business simulation games to enable students to acquire the key skills employers require EERN, Warwick, UK.

Rampasso, I. S., Anholon, R., Silva, D., Ordóñez, R. E. C., Quelhas, O. L. G., & De Santa-Eulalia, L. A. (2019). Developing in engineering students a critical analysis about sustainability in productive systems. International Journal of Sustainability in Higher Education. https://doi.org/10.1108/IJSHE-03-2018-0048

Raper, J. (2019). Project-Based Learning: add-on or the core of an engineering programme? EERN, Warwick, UK.

Reynolds, D., & Dacre, N. (2019). *Interdisciplinary Research Methodologies in Engineering Education Research*. Engineering Education Research Network, Coventry, UK. https://dx.doi.org/10.2139/ssrn.3812769

Romero, M., Usart, M., & Ott, M. (2015). Can serious games contribute to developing and sustaining 21st century skills? Games and culture, 10(2), 148-177. https://doi.org/10.1177%2F1555412014548919

Segalàs, J., Ferrer-Balas, D., & Mulder, K. F. (2010). What do engineering students learn in sustainability courses? The effect of the pedagogical approach. Journal of Cleaner Production, 18(3), 275-284. https://doi.org/10.1016/j.jclepro.2009.09.012

Segalàs, J., Mulder, K. F., & Ferrer-Balas, D. (2012). What do EESD "experts" think sustainability is? Which pedagogy is suitable to learn it? Results from interviews and Cmaps analysis gathered at EESD 2008. International Journal of Sustainability in Higher Education, 13(3), 293-304. https://doi.org/10.1108/14676371211242599

Smith, S. (2019). Exploration: Play in Practice—Innovation Through Play in the Postgraduate Curriculum. In The Power of Play in Higher Education (pp. 57-66). Springer.

Svanström, M., Gröndahl, F., Mulder, K. F., Segalàs, J., & Ferrer-Balas, D. (2012). How to educate engineers for/in sustainable development. International Journal of Sustainability in Higher Education. https://doi.org/10.1108/14676371211242535

Thürer, M., Tomašević, I., Stevenson, M., Qu, T., & Huisingh, D. (2018). A systematic review of the literature on integrating sustainability into engineering curricula. Journal of Cleaner Production, 181, 608-617. https://doi.org/10.1016/j.jclepro.2017.12.130

Watson, M. K., Lozano, R., Noyes, C., & Rodgers, M. (2013). Assessing curricula contribution to sustainability more holistically: Experiences from the integration of curricula assessment and students' perceptions at the Georgia Institute of Technology. Journal of Cleaner Production, 61, 106-116.
https://doi.org/10.1016/j.jclepro.2013.09.010

Yao, T., & Tite, C. (2019). An evaluation of student learning and reflection through the use of an engineering project management serious game EERN, Warwick, UK.

Yuan, X., & Zuo, J. (2013). A critical assessment of the Higher Education For Sustainable Development from students' perspectives–a Chinese study. Journal of Cleaner Production, 48, 108-115.
https://doi.org/10.1016/j.jclepro.2012.10.041

Zwikael, O., & Gonen, A. (2007). Project execution game (PEG): training towards managing unexpected events. Journal of European Industrial Training.
https://doi.org/10.1108/03090590710772668